# Wikipedia: A Challenger's Best Friend?

*Utilising Information-seeking Behaviour Patterns to Predict US Congressional Elections*

Hamza Salem[1] ✦ & Fabian Stephany ✦ ✧ [2]

✦ - Oxford Internet Institute, University of Oxford, UK
✧ - Humboldt Institute for Internet and Society, Berlin

**Working Paper**

**Abstract**

Election prediction has long been an evergreen in political science literature. Traditionally, such efforts included polling aggregates, economic indicators, partisan affiliation, and campaign effects to predict aggregate voting outcomes. With increasing secondary usage of online-generated data in social science, researchers have begun to consult metadata from widely used web-based platforms such as Facebook, Twitter, Google Trends and Wikipedia to calibrate forecasting models. Web-based platforms offer the means for voters to retrieve detailed campaign-related information, and for researchers to study the popularity of campaigns and public sentiment surrounding them. However, past contributions have often overlooked the interaction between conventional election variables and information-seeking behaviour patterns. In this work, we aim to unify traditional and novel methodology by considering how information retrieval differs between incumbent and challenger campaigns, as well as the effect of perceived candidate viability and media coverage on Wikipedia's predictive ability. In order to test our hypotheses, we use election data from United States Congressional (Senate and House) elections between 2016 and 2018. We demonstrate that Wikipedia data, as a proxy for information-seeking behaviour patterns, is particularly useful for predicting the success of well-funded challengers who are relatively less covered in the media. In general, our findings underline the importance of a mixed-data approach to predictive analytics in computational social science.

[1] For enquiries, comments, and suggestions please email Hamza Salem: hys228@nyu.edu
[2] Corresponding author: fabian.stephany@oii.ox.ac.uk

# Introduction

The prediction of electoral outcomes is not a new endeavour, and has remained a topic of public speculation in the United States since at least the 1880s (Rhode & Strumpf, 2004). Traditionally, professional election forecasting in the United States (US) has largely relied on direct polling (Hillygus, 2011). Some efforts have incorporated alternative methodologies that rely on indirect measures of public sentiment to bolster prediction literature. Stock market and gross domestic product performance (Abramowitz, 2008; Lewis-Beck & Rice, 1984), party related variables (Oppenheimer, Stimson, & Waterman, 1986), incumbency (Cover, 1977), and geographical levels of polarization and partisanship have often served as indirect behavioural predictors. For the purposes of our paper, we use the relatively novel indirect measure of public information-seeking behaviour patterns on Wikipedia to predict US Congressional Elections in 2016 and 2018.

Our indicator falls within the larger family of literature utilizing secondary measures of online generated data in political science. Extant scholarship in this field has been disproportionately dependent on social media data. Indeed, various modes of election prediction based on social media generated aggregates have been explored, with inconsistent success (Tumasjan et al., 2010; DiGrazia et al., 2014). However, potential biases in the population and temporal samples have sparked criticism of such conclusions (Jungherr, Jurgens & Schoen, 2012). In non-social media related research, Google search volume data have been considered as primary indicators, but have mostly demonstrated weak relationships with election outcomes (Lui, Metaxas, & Mustafaraj, 2011). Margolin et al. (2016) argue that the mixed results in election prediction using social media data are a result of platform features accentuating popularity rather than reflecting actual public interest in voting for specific candidates. The authors advocate for the use of online generated data that excludes reputational feedback mechanisms and highlights rational information seeking behaviour. We argue that Wikipedia pageviews largely satisfy this criteria.

Literature utilising Wikipedia pageviews and prediction span a number of topics including the spread of disease (Hickmann et al., 2015; Bardak and Tan, 2015) and stock market fluctuations (Moat et. al, 2013; Wei and Wang, 2016; Cergol and Omladic, 2015). Both foci highlight the importance of temporality and periodicity in the occurrence of relative spikes in interest. Similar to stock market prediction literature, we argue that Wikipedia pageviews are potentially a revealing indicator of political outcomes since they potentially shed light on voter decision-making processes. Throughout our analysis, we expand on this rational choice theory, arguing that the average voter will function similarly to a utility-maximizing rational investor.

Considering Wikipedia pageviews' potential for predicting electoral outcomes, there remains a surprisingly small number of peer-reviewed papers that use pageviews to predict elections.

Within extant scholarship, a similar argument of rational information-seeking behaviour is adopted with the hopes of testing Wikipedia's predictive usefulness. Throughout our paper, however, we build upon this literature by identifying the theoretical circumstances that may amplify or attenuate a potential relationship between pageviews and electoral outcomes. We attempt to unify traditional and novel data sources, within the context of US Senate and House elections, utilising both Wikipedia pageview data along with campaign indicators, including incumbency and challengership, perceived candidate viability, and television media coverage.

Our results largely confirm earlier findings that Wikipedia pageviews are indeed related to electoral outcomes. However, our analyses demonstrate that the predictive ability of pageviews remain highly nuanced. Pageviews are more likely to be indicative of positive electoral outcomes when the rationality behind the pageviews is explored. As such, we emphasize a number of moderating factors that may significantly alter the weight of each pageview in predicting elections, rather than relying solely on absolute volume of traffic.

In general, our study indicates that Wikipedia data, as a proxy for rational information seeking behaviour, is particularly useful in predicting the success of challenger, rather than incumbent campaigns. While incumbents' pages on Wikipedia are more likely to receive higher volumes of traffic, pageviews for challengers' are significantly more predictive of success, especially when the candidate is perceived to be viable. Our results also demonstrate that Wikipedia pageviews are most predictive in cases where candidates are underrepresented in public media outlets, and less useful when they are well-represented. These findings highlight the importance of distinguishing between exploratory and confirmatory information-seeking behaviours and their role in predicting electoral outcomes.

In the following sections, we present an overview of the literature on election forecasting using socially generated data, in particular of Wikipedia pageview statistics, from which we derive our hypotheses. We then outline our research design and follow with a description and discussion of our results. We conclude with limitations and potential steps for future research.

# Literature

## *Socially Generated Data*

A relatively novel method of measuring public opinion through passive means has been performed using the internet; particularly through the use of socially generated data. Such data arises as the result of interaction between individuals through the medium of the internet (Olshannikova et al., 2017). Theoretically, the ability of socially generated data to provide insight into public interest and opinion stems from the notion that the data are records of human behaviour on the internet. In the context of elections, capturing public opinion is contingent upon the internet's manifestation as an inclusive and discursive public sphere.

In mass media-oriented communicative structures, the parameters of public discussion remain limited as a result of the high barriers to entry into media dissemination (Habermas, 1991). In such cases, insight into public opinion is largely limited to survey and polling (Hillygus, 2011). The introduction and mass appeal of the internet, especially in its functionality as an unmediated means of communication, have shifted the dynamic of capturing public sentiment (Papacharissi, 2002; 2010). As a communicative space, as well as a high choice media environment, the internet affords researchers the ability to measure public interest and opinion formation by monitoring behavioural patterns as they relate to a certain subject. The shift from individual aspirational data in election polling to passive behavioural internet data informs models of human behaviour when considered in the aggregate. In the context of relevant literature, such models of human behaviour have been constructed to forecast a variety of events, including elections.

As we discussed in our introduction, much of the relevant prediction literature has been concerned with the use of social media, rather than Wikipedia pageviews, as a means to measure public interest for predictive purposes. Social networking platforms' mass adoption (Zickuhr, 2011) and the granularity of individual interactivity (Yu & Kak, 2012) present strong arguments for their inclusion in social science research. As indicators of human behaviour, social media data contain largely more revealing information relating to individual behaviour than Wikipedia pageviews.

However, we support our choice of indicator for a number of reasons. Primarily, we hold that the depth of knowledge captured by social media data invites a level of complexity and difficulty in expounding meaning and effect, especially as social media becomes more distinct in its own discourse and nuance (Gayo-Avello, 2012). Activity captured on social media platforms may be solely indicative of interactions on that platform rather than generalizable inferences of social phenomena (Cowls and Schroeder, 2015). Extant scholarship considering social media as a measure of public interest also falls prey to dependence on non-representative samples (Blank, 2016; Gayo-Avello, Metaxas, & Mustafaraj, 2011; Mislove et al., 2011). We believe Wikipedia usership to be significantly more representative of the general public both because it is a widely used resource (Schroeder & Taylor, 2015) and since it is used by a variety of individuals (Head & Eisenberg, 2010; Göbel & Munzert, 2018; Messner & South, 2011).

Wikipedia-specific indicators (including pageviews, page edits, and page creation) have been utilized to predict a plethora of topics (Stephany & Braesemann 2017), with a handful of peer-reviewed papers dedicated to the use of pageviews to forecast electoral outcomes. While there is growing scholarship around a number of other Wikipedia indicators, we concern our efforts solely with Wikipedia pageviews. In this regard, we maintain a similar scope that underpins the majority of pageview-related works which follow theories of rational choice and information-seeking behaviour. Wikipedia pageviews are assumed to offer a relatively accurate measure of public interest in gathering information on a particular subject.

### *Wikipedia Pageviews and Political Information-seeking*

Relevant scholarship has considered multiple metrics as indicators of political information-seeking patterns, including subscription to social networks (Himelboim, Hansen & Bowser, 2012) and television programming (Rubin, 1986). Regardless of means of exposure, however, information-seeking is predicated on an individual's rational intentionality in gaining further insight into a subject (Wilson, 1981). We rely on Wikipedia pageviews since it eliminates theoretical confounders found in other information-seeking processes, such as watching television or interacting on social media. Both the expectation of finding relevant targeted information and the presumed usefulness of that information are key in the motive of using Wikipedia to obtain data for a decision (Chung, 2012).

We argue that voters are more likely to seek information on relevant candidates about whom they are less-informed. Our reasoning follows a theory of optimal information-seeking presented by Ratchford (1982) in which consumers' decision-making (with regards to their search for relevant information) is contingent upon the cost of resources. Having already established Wikipedia's efficiency and wide appeal to a variety of audiences, it would not be unrealistic to consider its political usefulness. Indeed, a similar argument regarding cost-minimization in information retrieval, as well as a perception of unbiased content establishes Wikipedia as theoretically more efficient than reading multiple news articles and less biased than reading campaign-specific literature (Jordan, 2014).

The handful of peer-reviewed papers using Wikipedia pageviews to predict elections follow a similar line of argument of rational information-seeking behaviour patterns (see Table 1).

Yasseri and Bright (2013; 2016) find limited evidence that Wikipedia pageviews can independently predict vote share outcomes in a variety of election contexts. Most relevantly, the authors argue that information-seeking behaviour will be led by individuals seeking to vote differently from the past, possibly due to dissatisfaction with the incumbent (Yasseri & Bright, 2016). Smith and Gustafson (2017) consider Wikipedia pageviews as a proxy for interest in specific candidates in US Senate elections. The authors focus their efforts on using Wikipedia usage statistics to enhance predictions made by polling and fundamentals-based models. They introduce a synthesis of pageview statistics and polls as complementary measures, hypothesizing that relevant pageviews will predict unique variance in elections beyond that explained by polling aggregates. While neither set of authors find that Wikipedia can independently predict electoral outcomes, their results warrant further investigation into the instances in which Wikipedia might be most predictive.

**Table 1**. *Summary of relevant empirical literature studying the potential predictiveness of Wikipedia pageviews on electoral outcomes*

|  | Yasseri & Bright (2013) | Yasseri & Bright (2016) | Smith and Gustafson (2017) | Salem & Stephany (2020) |
|---|---|---|---|---|
| Dependent Variable | Vote Share | Vote Share | Vote Share | Vote Share & Absolute Outcome |
| *Wikipedia Indicators* | | | | |
| Page Views | x | x | x | x |
| *Campaign Variables* | | | | |
| Incumbency |  | x |  | x |
| Campaign Expenditure |  |  |  | x |
| Newness/Challenger Status |  | x |  | x |
| *Other Indicators* | | | | |
| Media Coverage |  | Print media |  | Television News |
| Google Trends | x |  |  |  |
| Poll results |  |  | x |  |
| *Study Details* | | | | |
| Years | 2010, 2013 | 2009, 2014 | 2008 - 2012 | 2016 - 2018 |
| Election Cycles | 3 | 2 | 3 | 2 |
| N | 16 (mixed) | 59 parties | <200 candidates | 1755 candidates |
| Geography & Setting | Multiple | Multiple | US Senate | House and Senate |

In building upon extant literature, we aim to contribute to the discussion of pageview predictivity by presenting a theoretical framework through which the nuances of information seeking patterns in elections can be studied. We develop a set of circumstances that may potentially affect Wikipedia's ability to predict congressional races. Namely, we attempt to uncover the effect of (1) candidate incumbency, (2) perceived viability, and (3) news media coverage on information-seeking patterns. By including these variables in separate prediction models, we hope to be able to both identify and explain their effect on Wikipedia's predictiveness. Further, while we do not incorporate polling data as Smith and Gusfatson (2017), we expand the dataset by using both US House and Senate elections, taking into account a variety of moderating variables that may alter the predictiveness of pageviews on election outcomes throughout the year. We also expand our dependent variable selection to include both vote share and absolute outcomes (*victory* and *defeat*)

# Approach and Hypotheses

*Rational Choice in Information-seeking Behaviour*

Throughout our approach, we assume a baseline level of social cognition in political decision-making. We rely on the theoretical rationality of voters in a democracy (Scott, 2000) for whom decision-making is contingent on having relevant information about political choices (Popkin, 1994). In general, we argue that utility-maximisation is paramount to individual vote choice (Downs, 1957), but that reaching an optimal choice requires information about candidate pay-offs (Lau & Redlawsk, 2006).

While rational, individuals are also limited in their decision-making ability by their own cognitive resources and the information available to them (Simon, 1957; Popkin, 1991). Average voters are therefore not capable, nor incentivized to seek out all necessary data regarding a particular subject, and instead are selective in their informational search (Fiske & Shelley, 1991; Stanovich, 2009) and will collect what data they deem to be sufficient to make informed choices (Scheufele, 2006).

We argue that the exposure to relevant information predates the act of information seeking, supporting the theory that active information-seeking is confirmatory, rather than exploratory (Lau & Redlawsk, 2006). This late stage step of targeted informational search in the cognitive process of decision-making is what is potentially best measured by Wikipedia pageviews. As such, an informational search may be indicative of a relatively predetermined decision to vote for a candidate.

Given Wikipedia's potential as a universal proxy of public information-seeking patterns, it would be logical to assume that candidates in a congressional race with relatively greater Wikipedia pageviews than their opponents are likely to receive a higher percentage of votes [3].

**H1:** *The proportion of pageviews a candidate receives compared to others in a race is a strong determinant of that candidate's vote share and absolute outcome in the race*

Further, we expect that undecided voters may remain undecided until late stages of the election, as they continue to receive passive political messaging (Valentino et al., 2004), at which point they are likely to seek targeted and confirmatory information. As more undecided voters seek information, a model that incorporates pageview proportionality will potentially be increasingly predictive of vote share, culminating in the highest predictiveness in the final weeks before an election.

### *Incumbency and Challengership*

Within the US congressional context, incumbency has been shown to significantly increase a candidate's likelihood of winning an election (Gelman & King, 1990). This incumbent advantage has been consistently attributed to a number of factors such as name recognition, party endorsement, financial support, and knowledge of relevant policy issues (Abramowitz, 1975). Incumbent campaign behaviour, perhaps owing to this advantage or to partisan loyalty (Peskowitz, 2019), often involves less concerted efforts of information gathering activities prior to elections when compared to relevant challengers (Hershey, 1973). A theoretical model that incorporates concepts of information-seeking behaviour would expect to see a mirror image of this disproportionality in the division of targeted information retrieval.

Our logic follows the aforementioned practical limitations of rational individuals such as cognitive miserliness, which will lead individuals to seek more information on new candidates rather than on those for whom they are likely to already have formulated an opinion. Further, we argue that increased rates of information-seeking regarding challengers, is indicative of dissatisfaction with the incumbent (Molyneux, 2004). Relevant theories support the notion that being dissatisfied with the status quo is likely to be a leading factor in research for alternative options (Johnson, 2002), and subsequently that this informational search will likely be of a targeted nature (Bimber, 1998). Therefore, the act of seeking information on challengers is more indicative of support than general curiosity, which may be a more prevalent phenomenon for incumbents.

---

[3] We do not assume that all pageviews translate to possible voters, and consider that a proportion of visits may be the result of non-standard editors (Göbel & Munzert, 2018). Notwithstanding such visitors, however, we believe an observation of pageviews and vote share across two congressional election cycles should be indicative of a potential relationship between pageviews and voting outcomes.

**H2:** *Wikipedia pageview statistics will be a stronger predictor of vote share and absolute outcome for challengers than incumbents.*

### *Perceived Viability*

While there is no singular measure of viability, scholarship often cites it as a set of factors that increase a candidate's chance of winning an election. Perceived viability, on the other hand, is the public perception and expectation that the candidate is capable of potentially winning a plurality of votes (Abramowitz, 1989). While electability may be best measured by opinion polling, perceived viability is often captured by fundraising success (Feigenbaum, 2008). We argue that campaign fundraising is a logical proxy for perceived viability in a US context since there is a long-standing correlation between campaign coffers and election outcomes, and because successful campaign fundraising efforts signal that the candidate has been able to raise enough funds from interested donors who equally believe in their electability. Historically, this has been especially true for challenger campaigns (Jacobson, 1978; Green & Krasno, 1988).

Relevant literature underscores the importance of this perception of candidate viability both in gaining voters (Abramowitz, 1988) and in the rate at which voters will seek information about the challenger (Utych & Kam, 2014). Once again, this aligns with general theories of information-seeking behaviour outlined above. As rational choice actors with a high concern for utility-maximization in their vote choice, and with limited resources, voters are unlikely to seek information regarding challenger candidates that they perceive to be nonviable contenders. We anticipate that perceived viability (as measured by campaign receipts) will lead to higher rates of targeted pageviews by potential voters of challenger candidates.

**H3:** *Higher proportions of campaign receipts will amplify the predictive effect of pageviews on vote share and absolute outcomes for challengers.*

Importantly, the hypothesis does not predict that larger campaign coffers will lead to more pageviews for challengers. Instead, our hypothesis lends itself to the discussion of the nuance in the predictivity of pageviews. Challengers perceived to be more viable will receive a larger proportion of pageviews from individuals that are more likely to vote for them. The distinction is made in order to avoid searching for direct effects of heavy campaign spending on increases in name recognition through television appearances, campaign rallies, or political advertisements.

*News Media Coverage*

An integral aspect of campaign information dissemination and mass political discourse in the US resides in television news coverage (Iyengar & Shanto, 2010). For decades, television news has been the most relevant source of campaign information (Chaffee & Kanihan, 1997). At the time of the 2016 and 2018 Congressional elections, television was the most widely used means of obtaining political news in the US (Mitchel et al., 2016; Shearer, 2018). In effect, news coverage of candidates may itself be a predictor of electoral outcomes, either causally in its capacity to elevate or diminish particular campaigns or by its very nature as a measure of campaign popularity.

Due to its relevance in public political discourse, media coverage is a logical predictor of public information-seeking. However, the relevance of congressional candidates within the larger political discourse is likely not equally distributed. Indeed, empirical evidence demonstrates disproportionately higher coverage of incumbents than challengers (Kahn, 1993; Schaffner, 2006; Green-Pedersen, 2017). Given their role in office, incumbents are more likely to be in the news when compared to relatively nascent challenger campaigners (Ordway, 2017). The extent to which news coverage may affect the rates of relevant information-seeking of incumbent campaigns, however, is potentially less dramatic than that of challengers who remain relatively less well known and are more likely to be researched.

**H4(a):** *News media coverage is a stronger predictor of pageviews for challengers than for incumbents throughout an election year.*

Further, news coverage of a candidate may have a moderating effect on the relationship between pageviews and vote share. While for some audiences, news coverage may satisfy the conditions required for vote-choice (Iyengar, 1990), it may encourage others to seek further information about candidates. Considering the theories of limited rationality and of convenience in informational search (Connaway, Dickey, & Radford, 2011), however, the paper predicts that an increase in news coverage will reduce the effect of pageviews on vote share. While more frequent news coverage of candidates will lead to higher pageview totals, the likelihood that these page visits reflect potential interest in a candidate decreases.

**H4(b):** *As news coverage of a candidate increases, the effect of Wikipedia pageviews on vote share for that candidate will decrease*

# Methodology and Data

To assess the predictiveness of pageviews on election outcomes, we collected daily Wikipedia pageviews for all unique US congressional candidates who ran in either one or both of the two separate elections in 2016 and 2018 (N=1755). Throughout the paper, we divide variable data along Senate and House lines [4].

## *Dependent Variables*

Our analyses rely on the proportion of Wikipedia pageview statistics of race-specific candidates in the year leading up to a general election. We consider both the relative *vote share* and *absolute outcomes* of each race.

*Vote share and absolute outcome* – US Senate and House of Representatives election data from two general elections (2016 and 2018) were separately collected from a dataset compiled by the MIT Election Data Lab. From the available data, it was possible to determine the vote share of each candidate and ascertain the winner of each race. We incorporate both dependent variables since the US electoral tradition follows a system of plurality voting and may include more than two candidates. As such, the proportion of votes needed to secure a victory are not representative of any preset threshold. More importantly, however, there is a common bimodal distribution of candidates across most measurements such as funding, pageviews, news coverage, and challenger and incumbent vote share (see Figure 1). Given these realities, while vote share remains the major dependent variable throughout the paper, absolute outcomes are considered where most necessary and useful.

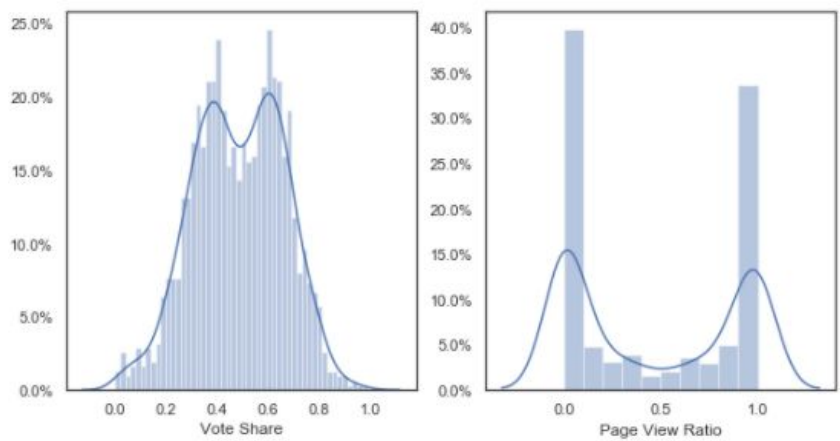

**Figure 1:** *Distribution of vote share and pageview ratio among House candidates*

---

[4] The United States Congress is divided into 100 elected senators, two from each of the 50 US States, and 441 elected House Representatives each representing a unique district.

*Independent Variables*

*Wikipedia Pageview Data* – Historical pageview data consists of the number of times an existing article was read in daily, monthly, and yearly formats. We utilized the MediaWiki API to collect daily pageview information for each candidate who had a dedicated page on Wikipedia over the course of the election year which began 365 days before the election day [5].

*Incumbency and Challenger Status* – To determine the status of each candidate as an incumbent or challenger we ran their name against historical data from the MIT Election Data Lab dataset.

*Campaign Finance Data (Viability)* – Following the theory outlined in our approach, campaign receipts were used as a proxy for a candidate's perceived viability. These data were obtained from the official Federal Election Commission (FEC) database[6].

*Media Coverage* – We used the GDELT TV API to obtain broadcast news collected by the Internet Archive's Television News Archive[7]. The API was automatically set to search the dataset for instances in which a candidate was mentioned by a predefined list of television channels. The parameters of the search included the candidate's first and last name being mentioned consecutively across a number of partisan and local news channels [8]. The inclusion of multiple channels, while proving to be logistically difficult, was necessary to ensure representativeness in coverage.

*Transformations*

*Time-related transformations* – We reorganized Wikipedia and media coverage data from daily to average weekly data to avoid overweighting singular events that may have led to spikes in interest in candidates. Since two different years were taken into consideration, original dates were renumbered from 0 to 51, with the latter being the week of election day. This standardization was performed to avoid splitting the data into two separate datasets for 2016 and 2018.

*Proportional Transformations* – To calibrate the dependent variable of vote share (which ranged from 0 to 1) with independent variable values, we performed a series of transformations of candidate-specific values from raw data to ratios. Each candidate's value of Wikipedia pageviews, campaign receipts, and news media mention totals were recalculated to become proportions of the state or district race totals. In place of raw pageviews on certain weeks or specific amounts of campaign funding in US dollars, all candidates now had a variable ratio that was indicative of their performance relative to their race specific contenders [9]. This method also allowed us to control for differences in state and district specific populations.

---

[5] The election day always falls on the first Tuesday after November 1 of even years
[6] The FEC is the regulatory body which monitors and enforces campaign financing laws and regulations in the US
[7] The News Archive is a large database of consecutive 15-second, captioned television news broadcasting
[8] We included CNN, MSNBC, Fox News, and local CBS affiliates

*Binary Transformations* – A similar method of transformation was undertaken to calibrate the ratio values with the binary outcome variable of victory and defeat. In this case, the candidate with the highest value referring to any particular ratio in a specific time frame was awarded a 1. They otherwise received a 0.

### *Measurements*

Where the dependent variable was continuous (i.e. vote share), OLS regression calculations were used to obtain the proportion of the variance in vote share explained by Wikipedia pageview ratios alone. In testing the independent variable according to our theoretical approach, moderating variables were introduced as interaction terms to measure the change in both $R^2$ and the *beta* coefficient .

In the secondary measurement of absolute outcomes, the categorical variable of pageview ratio was used as the independent variable in a binomial logistic regression model, in the form:

$$ln(P/1 - P) = \beta_0 + \beta_1(X_1)$$

where *P* is the probability of a particular outcome of a candidate's campaign, $\beta_0$ is the intercept, $\beta_1$ is the main coefficient, and $X_1$ is the pageview ratio, or the transformed categorical value (pageview outcome).

In the final step, a logistic regression model that includes a majority of the independent variables was created to test both the accuracy of the model and the importance of each variable under different conditions. The pageview outcome variable was then included to test whether it added to the model's accuracy and significantly changed relevant probability values for challengers. The binomial logistic function was of the form:

$$ln(P(ChallengerVictory))/1 - P(ChallengerVictory)) = \beta_0 + \beta_1(PageviewOutcome) + \beta_2(PerceivedViabilityOutcome) + \beta_3(OpenSeat/Incumbent)$$

---

[9] For instance, a candidate with 6,500 pageviews at the end of the year in a race where all candidate pageviews totaled 78,000 views now had a pageview ratio of 6,500/78,000 = 0.08, or 8 percent.

# Results and Discussion

Throughout the results, House and Senate data are measured separately and focus is placed on presenting House data, as there are substantially more House candidates (N=1590) than Senate candidates (N=165). In cases where Senate results are not present in the main body of the following sections, they are included in the appendix.

### Hypothesis 1: *Pageviews, vote share, and absolute outcomes*

We begin by testing the ability of pageview statistics to predict vote share independently. To do so, we consider the transformed race-specific pageview ratios, such that both dependent (vote share) and independent variables (pageview ratio) range from 0 to 1. When measuring the variance in vote share, pageview ratio can explain a significant amount of variance for both House ($R^{2(adj)} = .46$, $p<.001$) and Senate candidates ($R^{2(adj)} = .50$, $p<.001$) (see Appendix A). Our first hypothesis, which stated that a candidate's pageview ratio is predictive of vote share, is supported[10].

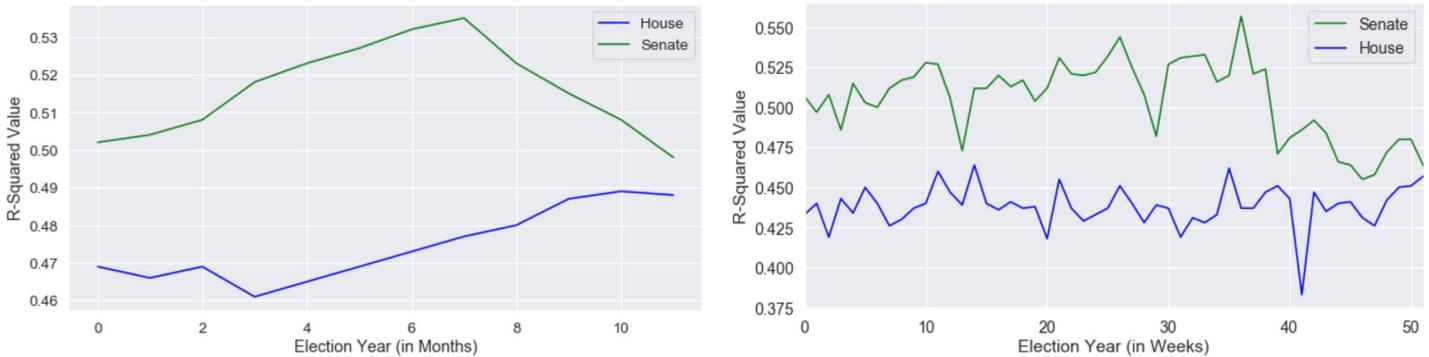

*Note*: *p<.001 across each observation throughout the election*

**Figure 2.** *The monthly cumulative rate (above) and weekly non-cumulative pageview rate (below) predicting variance in vote share throughout the election year*

Since our initial research question included an emphasis on the temporality of information seeking patterns, we consider the variance in vote share explained by pageviews throughout the year. We also make a distinction between the cumulative and non-cumulative values of pageviews at certain intervals during the year. As can be seen in Figure 2, Wikipedia pageviews' ability to explain variance in vote share for House and Senate candidates remains fairly stable throughout the year.

---

[10] Importantly, we found that the distribution of candidates across pageview ratios is bimodal (see Appendix B), with a large contingent located near the 0 and 1 regions of the plot. This is due to the reality that competitive race-specific candidates with large proportions of pageview ratios are guaranteed to be met with opposing candidates with small proportions of pageviews.

Considering that the variance explained does not widely differ across the year, we use the election day cumulative pageview ratio as the main independent variable throughout the remainder of the paper[11]. Interestingly, our expectation that Wikipedia would be most predictive towards the end of the election cycle was not met. However, Wikipedia pageviews do provide early insight into vote share election outcomes.

We now move on to testing Wikipedia's ability to predict House election outcomes in absolute terms, or the probability of a candidate's victory or defeat. We set out the value of such a measurement in a plurality voting system in our methods section. The pageview ratio is used in two different logistic regression analyses. In the first instance, the independent variable of pageview ratio is transformed into a categorical variable, in which a candidate receives a 1 if they maintained a higher pageview ratio than all other opponents in the race and 0 if they did not.

When running a logistic regression for races in which some candidates did not not have a page, the model can predict the winner of the election with 0.97 accuracy. When a candidate had no Wikipedia page, and their opponent did, the probability of that candidate winning the election was 2 percent (see Appendix C). This low ratio demonstrates the importance of having a dedicated Wikipedia page throughout the election year. In referring back to the theory, the eventuality that a candidate did not own a page may itself be indicative of low public interest levels (Margolin, et al., 2016). However, in reality, along with polling and other measures such as media coverage and candidate name recognition, the defeat of a candidate without a dedicated Wikipedia page would not be a surprising result.

In more competitive races in which both candidates had a dedicated Wikipedia page, a model that takes into account a binary value of pageview outcome, has an accuracy of .70 (See Appendix D). Following the same logic above, the probability of winning a race if a candidate had proportionally more pageviews than all other opponents is 68 percent. If the pageview ratio remained a continuous variable, the model is equally accurate. At a pageview ratio value of 0.5, the probability of winning the race is 51 percent. At a value of .6, the probability of winning is 57 percent. Thus, an increase of 10 percent in pageview ratio increases the probability of winning by 6 percent.

Our results in testing the first hypothesis demonstrate that pageview ratios can explain a significant proportion of variance in vote share in both House and Senate elections.

---

[11] For House candidates, $R^{2(adj)} = .46$, $p<.001$, and Senate candidates $R^{2(adj)} = .50$, $p<.001$

Theoretically, the results demonstrate that information-seeking patterns measured by pageviews correlate with election outcomes. Further, information-seeking appears to take place throughout the year at a relatively stable pace. These results do not consider a number of moderating and possibly explanatory variables. As such, there is a possibility that these results are epiphenomenal, and any insight offered by them can be gained by considering other variables. This potentiality is tested in the following hypotheses.

### Hypothesis 2: *Challengers and Incumbents*

**Table 3**. *Regression Analysis Predicting Vote Share from Pageview Ratio (Model 1), Challenger Status (Models 2, 3 and 4) and Race Type (Model 5)*

| | \multicolumn{5}{c}{*Dependent variable:*} |
|---|---|---|---|---|---|
| | \multicolumn{5}{c}{House Candidate Vote Share} |
| **Measure** | (1) | (2) | (3) | (4) | (5) |
| Pageview Ratio | .273*** (<.01) | | .151*** (<.01) | .053*** (.012) | .578*** (<.01) |
| Challenger Status | | -.255*** (<.01) | -.168*** (<.01) | -.259*** (.012) | -.247*** (<.01) |
| Pageview * Challenger Status | | | | .147*** (.012) | |
| Open Seat Race | | | | | .04*** (.01) |
| Pageview * Open Seat | | | | | .15*** (.01) |
| Constant | 0.355*** (<.01) | .635*** (<.01) | .512*** (<.01) | .592*** (.01) | .578*** (<.01) |
| Observations | 1590 | 1590 | 1590 | 1590 | 1590 |
| $R^2$ | 0.463*** | 0.508*** | 0.590*** | 0.607*** | 0.637*** |
| Adjusted $R^2$ | 0.463*** | 0.508*** | 0.590*** | 0.607*** | 0.636*** |

*p<0.05, **p<0.01, ***p<0.001

Our second hypothesis stated that pageview ratio statistics would be more predictive of vote share for challengers than for incumbents. We reasoned that demand for information about incumbents, who are likely to be more well known than their opponents, should generally be lower owing to the cognitive and resource limitations that would reduce the proportion of search for already established information. In order to test this hypothesis, two scenarios in which challengers are present should be taken into account. In the first, and more likely scenario, challengers run against incumbents. In the second, challengers run against other non-incumbents in an *open seat* race. We specify these scenarios in our model outlined in Table 3.

US congressional races in 2016 and 2018, and indeed for the majority of races in modern US history, are won by incumbents (see Appendix E). In Table 3, the introduction of the binary variable of challenger status, which denotes that the candidate was a non-incumbent, demonstrates the negative effect that being a challenger has on vote share. Taking into account this binary variable as an independent term in model (3) demonstrates that the amount of variance in vote share explained increases by 0.13, ($R^{2(adj)} = .59$, $p < .001$) when compared to the baseline model.

While challenger status has a negative effect on vote share, it has a positive effect on the relationship between pageviews and vote share, which can be seen by the positive interaction term coefficient ($b = .15$, $p < .001$). Similarly, the addition of the binary variable of open-seat race status also increases the effect of pageview ratio in predicting vote share for challengers. As such, these results support the hypothesis that pageview statistics are more predictive of election outcomes for challengers.

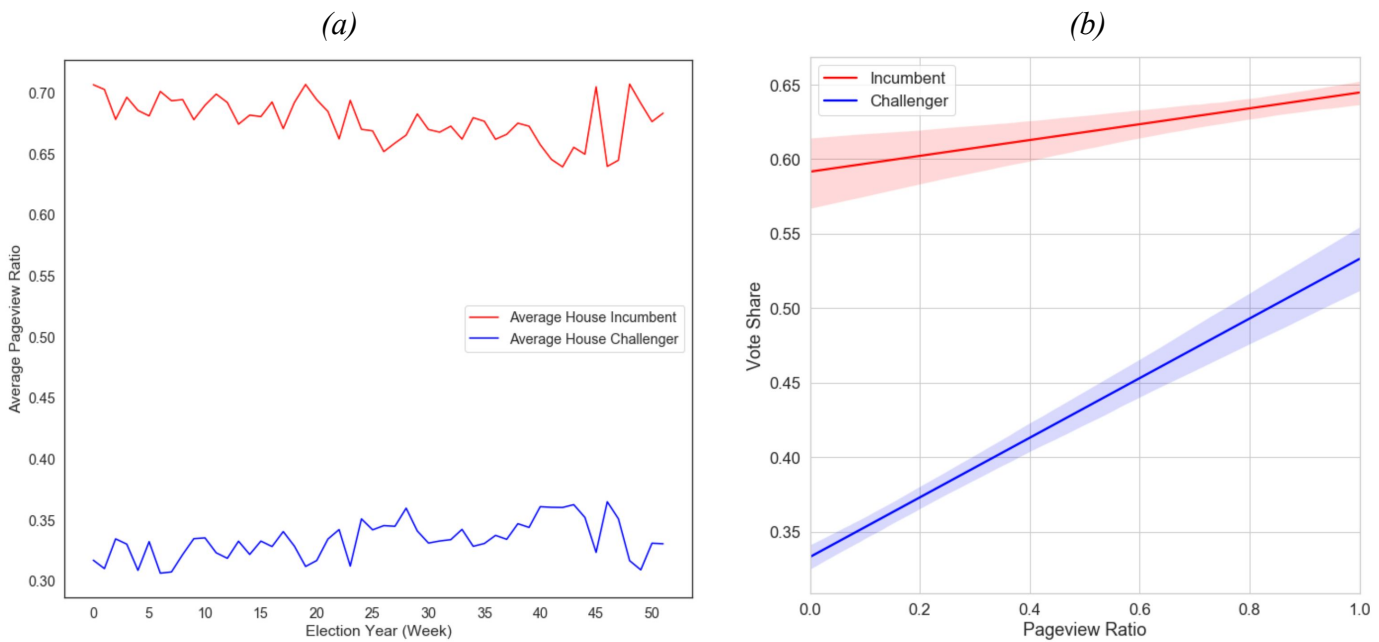

**Figure 3**. *Average pageview ratio of incumbents and challengers throughout election cycle (a) and relationship between pageview ratio and candidate types (b)*

Interestingly, the average House incumbent maintained a significantly higher pageview ratio than the average challenger throughout the year (Figure 3.a). This finding challenges the notion that non-incumbents enjoy more attention on Wikipedia, even when considering that incumbents are more widely known and would theoretically be researched less often. However, this average is not indicative of the predictive effect of pageviews on the different types of challengers. Instead, Figure 3.b, which presents the slope between pageview ratio and vote share for different candidate types, demonstrates that an increase in the pageview ratio of 10 percent for incumbents corresponds to an increase of 0.5 percent in vote share. For challengers, an equal increase in pageview ratio correlates with a 2 percent increase in vote share.

It would be difficult to determine the exact cause of such results without a controlled experimental approach. However, the relevant theory we present in our approach is quite robust in providing reasons for the usefulness of Wikipedia metadata for challengers who are, in a majority of cases, less likely to enjoy public recognition. Thus, a visit to their dedicated page is more indicative of a potential voter when compared to incumbent page visits. Another potential factor present in these observations is dissatisfaction with the incumbent. Voters may seek more information about challengers in cases where they are dissatisfied with the incumbent, which may help explain the larger coefficient size for challengers than incumbents. As such, the effect of visiting a challenger's page is four times larger than visiting an incumbent's.

### Hypothesis 3: *Perceived Viability*

As we discussed in our approach, relevant theory with regards to perceived candidate viability is highly related to the candidate incumbency status. Following this theory, we hypothesized that a challenger's viability ratio will have a larger positive effect on the relationship between pageviews and electoral outcomes. In Table 4, the addition of an interaction between viability ratio, challenger status, and view ratio in model (6) increases the proportion of variance explained such that $R^{2(adj)} = .71$, $p < .001$. The coefficient of the interaction is also positive and significant ($b = .12$, $p<0.01$), demonstrating that challengers who are (or are publicly perceived to be) more viable enjoy a stronger relationship between pageview ratio and vote share. As such, our hypothesis is supported.

**Table 4.** *Regression Analysis Predicting Vote Share from Pageview Ratio (Model 1) and Viability Ratio (Models 2 - 6)*

| Measure | \multicolumn{6}{c}{*Dependent variable:* House Candidate Vote Share} | | | | | |
|---|---|---|---|---|---|---|
| | (1) | (2) | (3) | (4) | (5) | (6) |
| Pageview Ratio | .273*** (<.01) | | .15*** (<.01) | .184*** (.02) | | .1** (.03) |
| Viability Ratio | | .321*** (<.03) | .23*** (<.01) | .263*** (.01) | .132*** (.01) | .18*** (.04) |
| Challenger Status | | | | | -.219*** (.01) | -.159*** (.03) |
| Pageview Ratio * Viability Ratio | | | | -.069*** (.02) | | |
| Viability Ratio * Challenger Status | | | | | .127*** (.02) | |
| Viability Ratio * Challenger Status * Pageview Ratio | | | | | | .122** (.05) |
| Constant | .355*** (<.01) | .33*** (.01) | .303*** (<.01) | .294*** (<.01) | .531*** (<.01) | .458*** (.03) |
| Observations | 1590 | 1590 | 1590 | 1590 | 1590 | 1590 |
| $R^2$ | .463*** | .563*** | .658*** | .66*** | .673*** | .709*** |
| Adjusted $R^2$ | .463*** | .563*** | .658*** | .66*** | .671*** | .708*** |

*p<0.05, **p<0.01, ***p<0.001

The addition of an interaction term between viability and pageview ratio without challenger status distinction in model (4) is negative (*b*=.-07, *p<0.001)*. Indeed, when accounting for incumbency status, an interaction between viability and pageview ratio has an even larger negative effect size (see Appendix F). A potential explanation involves the effect of campaign receipts, and subsequently campaign spending, on increasing the sources of information for congressional candidates through public facing events and television advertisements. This may potentially lead to a higher level of public interest in relatively less known challenger candidates and increase information-seeking concerning them.

In cases of incumbents, negative ramifications of a candidate being perceived to be well-funded when that candidate currently holds office have been observed historically (Coates, 1998; Moon, 2006). While incumbents enjoy more attention on Wikipedia relative to challengers, the predictiveness of pageviews is weaker on average than for challengers running in the same race. When incumbents have higher perceived viability ratios the ability of pageviews to predict vote share for those incumbents decreases further.

Hypothesis 4: *Media Coverage*

Our final hypothesis predicted that news media coverage ratios will be a stronger predictor of pageview ratios for challengers than for incumbents. This aligned with our underlying theoretical narrative that incumbents are less likely to be researched than challengers. We expected that the mention of relatively new challengers is likely to lead more public inquiry into the candidate when compared to well-known incumbents. Figure 6 (below) confirms this hypothesis.

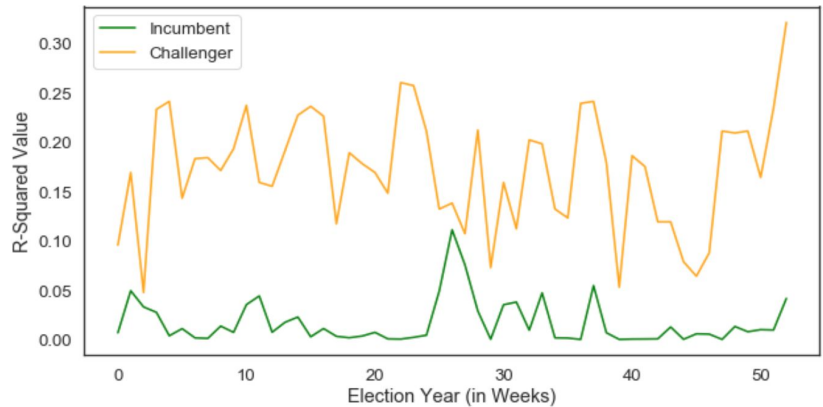

**Figure 4**. *Proportion of variance in pageview ratio predicted by media coverage ratio for incumbents and challengers*

While television coverage of incumbents is higher on average than challengers throughout the election year (see Appendix G), it is more predictive of pageview ratios of challengers (see Figure 4). However, what is most interesting in the illustration is the limited proportion of variance in pageview ratio explained by news coverage, which at it its highest level (the final week of the election cycle) explains only 33 percent ($R^{2(adj)} = .30$, $p<.001$). This supports the potentiality that pageviews are not epiphenomenal. Further, our results demonstrate that Wikipedia pageviews are potentially more predictive of vote share than news coverage. In Table 6 (below) news coverage ratio (model 2) explains 20 percent ($R^{2(adj)} = .30$, $p<.001$) less variance in vote share than model (1) which features pageview ratio as the sole independent variable ($R^{2(adj)} = .50$, $p<.001$).

**Table 6.** *Regression Analysis Predicting Vote Share from Pageview Ratio (Model 1), News Ratio (Models 2 - 4)*

|  | Dependent variable: | | | |
| --- | --- | --- | --- | --- |
|  | Senate Candidate Vote Share | | | |
| **Measure** | (1) | (2) | (3) | (4) |
| Pageview Ratio | .368*** (.03) |  | .308*** (.04) | .379*** (.04) |
| News Coverage Ratio |  | .237*** (.03) | .084*** (.03) | .181*** (.05) |
| Pageview Ratio * News Ratio |  |  |  | -.187** (.07) |
| Constant | .248*** (.02) | .31*** (.02) | .241*** (.02) | .222*** (.02) |
| Observations | 165 | 165 | 165 | 165 |
| $R^2$ | .498*** | .297*** | .522*** | .541*** |
| Adjusted $R^2$ | .495*** | .293*** | .516*** | .532*** |

*p<0.05, **p<0.01, ***p<0.001

In Table 6, model (3) which includes both pageview and news ratios, the proportion of variance in vote share that is explained by the model does not increase significantly. This may indicate that the effect of news coverage on predicting variance in vote share is already explained by Wikipedia pageviews. Model (4), which introduces the interaction term between pageview and news ratios, similarly explains only a slightly higher proportion of variance ($R^{2(adj)} = .54$ $p<.001$). Interestingly, the interaction term's coefficient is negative ($b = -.19, p<0.01$), which suggests that higher levels of coverage corresponds with lower levels of pageview predictiveness.

While it would be difficult to claim a definitive reason for this observation, we argue that it may potentially be the result of news coverage leading to general, rather than confirmatory inquiry into candidates running in elections. Pageviews may be indicative of curiosity after exposure rather than interest in voting for the candidate. The final hypothesis, which predicted that higher ratios of news coverage would reduce the relevance of Wikipedia in explaining vote share, is supported

*Predicting Probabilities*

Given the disproportionate distribution of candidates across most variables as well as the reality of the US plurality voting system, binary absolute outcomes may be a useful secondary dependent variable to consider prior to concluding our efforts. We believe that the potential ineffectiveness of vote share in predicting winners, owing to the lowered threshold in races in which multiple candidates run and divide the vote share, may be corrected by considering such absolute outcomes.

A binomial regression incorporating the same set of variables discussed throughout the paper produces a model that has an accuracy rate of 94 percent (See Appendix H). The addition of pageview ratio to this model does not add to the model's accuracy. However, incorporating the transformed variable of pageview outcome does offer some insight into the importance of the variable in such a model.

**Table 7**. *All possible variable value combinations and the corresponding probability of challenger victory*

| Primary Candidate | Opponent Type | Candidate Viability Status | Challenger Pageview Outcome | Probability of Challenger Victory | Addition of Pageview Outcome P(%) |
|---|---|---|---|---|---|
| Challenger | Incumbent | Less viable | Less pageviews | 2% | + 6 |
| | | | More pageviews | 8% | |
| | | More viable | Less pageviews | 18% | + 24 |
| | | | More pageviews | 42% | |
| | Open Seat | Less viable | Less pageviews | 16% | + 23 |
| | | | More pageviews | 39% | |
| | | More viable | Less pageviews | 64% | + 21 |
| | | | More pageviews | 85% | |

The results in Table 7 demonstrate the value of having a higher proportion of Wikipedia pageviews than an opponent for challengers. For instance, where the challenger has financially out-raised an incumbent, the probability of challenger victory is 18 percent when that candidate does not have a higher proportion of pageviews than the incumbent. However, the additional value of meeting the pageview victory condition increases the probability of winning by 24 percent (*P*(*win*) = .42). When running in open seat elections, both viable and non-viable challengers have an additional ~22 percent chance of winning if they maintain a higher proportion of pageviews.

It is important to note that this method overlooks a substantial amount of potential variance in the data and the probability value may not be the most accurate measure of candidate success. However, given the model's accuracy rate and the concentration of data points in the ends of the spectrum in a majority of variable ratio distributions, these observations produce insight that would otherwise be lost when considering change in variance in vote share only.

## Conclusion

Throughout this paper, we explore both the theoretical and empirical aspects of Wikipedia pageview predictiveness of electoral outcomes. We argue that pageviews largely reflect rational information-seeking behaviour patterns, and thus are indicative of general public sentiment. However, we produce a number of crucial nuances and instances in which the predictive effect of pageview data may attenuate and amplify such a relationship.

Generally, our results demonstrate that Wikipedia pageviews are most predictive of challenger campaigns; in particular, those that are perceived to be viable. The effect of pageviews was four times stronger for challengers than for incumbents in predicting vote share, even as incumbents maintained higher pageview ratios on average. Additionally, we find that media coverage of candidates is more predictive of challenger pageviews, but that the relationship remains quite weak throughout the election cycle. As media coverage of a candidate increases, pageview ratios generally lose relevance as predictive markers for vote share. This phenomenon remains true even as the volume of traffic to pageviews grows with higher rates of media coverage.

We conclude that our initial premise is likely to be true; Wikipedia's role as an information provider, and as a source of predictive markers, is strongest when conventional information sources are nonsignificant. Further, our approach allows us to delineate between Wikipedia pageviews that are confirmatory, and therefore predictive, and those that are exploratory and indicative of general curiosity.

We recognize, however, that our work comes with limitations. First, we focus our analysis on US Senate and House elections. The nuances of the American electoral context reduce the generalizability of our results, since many systems do not rely on plurality voting mechanisms.

Second, while we underline the relevance of Wikipedia data for less popular candidates, it might be useful to cross-reference our indicators with other information-seeking proxies such as search engine queries to include data on candidates that have no dedicated Wikipedia pages. Finally, with regard to perceived viability, the addition of individual donors, rather than campaign receipts may have been a theoretically more meaningful indication of public support and candidate viability, however, these data were unavailable.

In its totality, this paper underlines the importance of interacting conventional campaign characteristics with online generated features to elucidate motive and meaning in datasets with low-dimensionality. Our unification of traditional prediction features and novel online generated markers highlights the relevance of a mixed-data approach to predictive analytics in computational social science. As the use of socially generated data continues to grow in fields of social and political prediction, scholars should reinforce the use of mixed-data approaches in the effort to understand the nature and consequence of online human behaviour.

# Appendix

**Appendix A.** *Regression Analysis Predicting Vote Share from Pageview Ratio*

| Measure | N | B | SE B | F-Statistic | DF | $R^2$ | $R^{2(adj)}$ |
|---|---|---|---|---|---|---|---|
| **Pageview Ratio** | | | | | | | |
| Senate Candidate Ratio | 165 | 0.37*** | .03 | 187.5 | 1 | 0.50*** | 0.50*** |
| House Candidate Ratio | 1590 | 0.27*** | <.01 | 1447 | 1 | 0.46*** | 0.46*** |

***p<.001

*Note: the value of pageview ratio for each candidate type is calculated as the cumulative total on the week of the election.*

**Appendix B.** View Ratio predicting Vote share for House (left) and Senate (right)

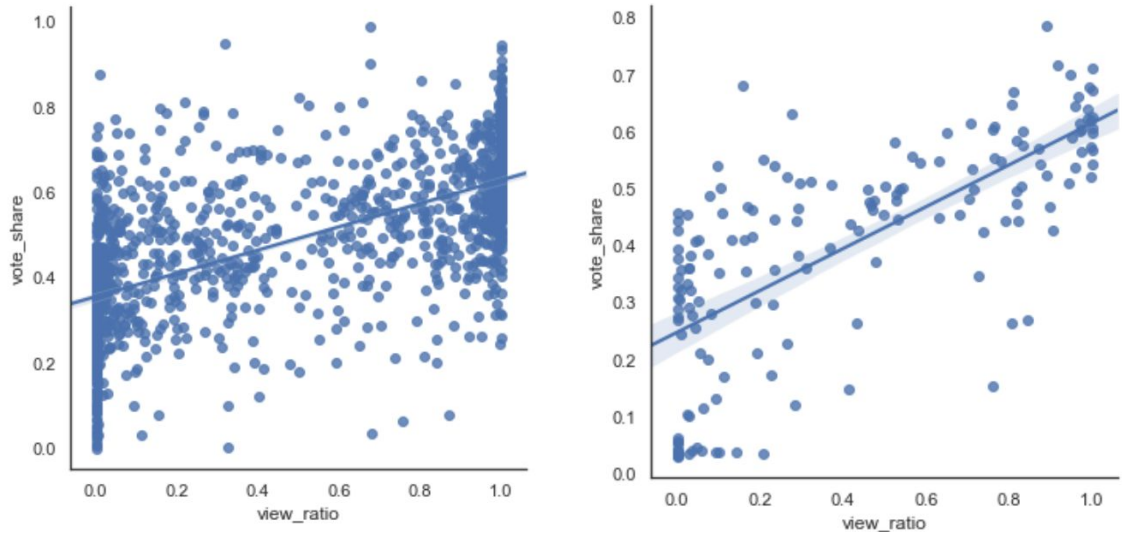

**Appendix C.** *Logistic Regression Analysis predicting odds of winning in cases where opponents did not both have a Wikipedia page*

```
                        Logit Regression Results
==============================================================================
Dep. Variable:                 winLose   No. Observations:                  782
Model:                           Logit   Df Residuals:                      780
Method:                            MLE   Df Model:                            1
Date:                 Mon, 06 Jul 2020   Pseudo R-squ.:                  0.8285
Time:                         10:20:55   Log-Likelihood:                -92.902
converged:                        True   LL-Null:                       -541.67
Covariance Type:             nonrobust   LLR p-value:                 3.361e-197
==============================================================================
                 coef    std err          z      P>|z|      [0.025      0.975]
------------------------------------------------------------------------------
Intercept     -3.7740      0.337    -11.194      0.000      -4.435      -3.113
view_win       7.2896      0.455     16.012      0.000       6.397       8.182
==============================================================================
```

**Appendix D:** *Logistic Regression Analysis predicting odds of winning in cases where opponents both have a Wikipedia page and Page Outcome is binary (top) and Pageview Ratio is continuous (bottom)*

```
                        Logit Regression Results
==============================================================================
Dep. Variable:                 winLose   No. Observations:                  808
Model:                           Logit   Df Residuals:                      806
Method:                            MLE   Df Model:                            1
Date:                 Mon, 06 Jul 2020   Pseudo R-squ.:                  0.1119
Time:                         10:35:30   Log-Likelihood:                -497.41
converged:                        True   LL-Null:                       -560.06
Covariance Type:             nonrobust   LLR p-value:                  4.377e-29
==============================================================================
                 coef    std err          z      P>|z|      [0.025      0.975]
------------------------------------------------------------------------------
Intercept     -0.8226      0.108     -7.604      0.000      -1.035      -0.611
view_win       1.6407      0.153     10.743      0.000       1.341       1.940
==============================================================================
```

```
                        Logit Regression Results
==============================================================================
Dep. Variable:                 winLose   No. Observations:                  808
Model:                           Logit   Df Residuals:                      806
Method:                            MLE   Df Model:                            1
Date:                 Mon, 06 Jul 2020   Pseudo R-squ.:                  0.1255
Time:                         10:37:24   Log-Likelihood:                -489.75
converged:                        True   LL-Null:                       -560.06
Covariance Type:             nonrobust   LLR p-value:                  1.948e-32
==============================================================================
                 coef    std err          z      P>|z|      [0.025      0.975]
------------------------------------------------------------------------------
Intercept     -1.1237      0.127     -8.878      0.000      -1.372      -0.876
view_ratio     2.3281      0.209     11.129      0.000       1.918       2.738
==============================================================================
```

**Appendix E.** *The total number of successful and unsuccessful congressional challengers and*

*incumbents in 2016 and 2018 (combined)*

|  | Total Candidates | Open Seat Races | Challenger Victory | Challenger Defeat | Incumbent Victory | Incumbent Defeat |
|---|---|---|---|---|---|---|
| *Senate* | 165 | 15 | 22 | 90 | 46 | 7 |
| *House* | 1590 | 138 | 176 | 769 | 607 | 38 |

**Appendix F.** *Pageview ratio, incumbency and viability ratio (demonstrating the inverse effect of viability on incumbents) for House (top) and Senate (bottom)*

```
                            OLS Regression Results
==============================================================================
Dep. Variable:             vote_share   R-squared:                       0.709
Model:                            OLS   Adj. R-squared:                  0.708
Method:                 Least Squares   F-statistic:                     550.5
Date:                Mon, 06 Jul 2020   Prob (F-statistic):               0.00
Time:                        15:25:52   Log-Likelihood:                 1491.6
No. Observations:                1590   AIC:                            -2967.
Df Residuals:                    1582   BIC:                            -2924.
Df Model:                           7
Covariance Type:            nonrobust
========================================================================================
                                           coef    std err          t      P>|t|      [0.025      0.975]
----------------------------------------------------------------------------------------
Intercept                                0.2986      0.004     66.535      0.000       0.290       0.307
view_ratio                               0.1065      0.013      8.049      0.000       0.081       0.132
receipt_ratio                            0.1920      0.013     14.917      0.000       0.167       0.217
view_ratio:receipt_ratio                 0.0525      0.024      2.145      0.032       0.004       0.100
incumbency                               0.1591      0.029      5.555      0.000       0.103       0.215
view_ratio:incumbency                   -0.0109      0.036     -0.300      0.764      -0.082       0.060
receipt_ratio:incumbency                -0.0080      0.037     -0.216      0.829      -0.081       0.065
view_ratio:receipt_ratio:incumbency     -0.1227      0.048     -2.577      0.010      -0.216      -0.029
==============================================================================
Omnibus:                       33.402   Durbin-Watson:                   2.529
Prob(Omnibus):                  0.000   Jarque-Bera (JB):               62.021
Skew:                          -0.119   Prob(JB):                     3.41e-14
Kurtosis:                       3.938   Cond. No.                         49.1
==============================================================================
```

```
                            OLS Regression Results
==============================================================================
Dep. Variable:             vote_share   R-squared:                       0.661
Model:                            OLS   Adj. R-squared:                  0.652
Method:                 Least Squares   F-statistic:                     77.84
Date:                Thu, 02 Jul 2020   Prob (F-statistic):           1.55e-36
Time:                        14:19:14   Log-Likelihood:                 128.11
No. Observations:                 165   AIC:                            -246.2
Df Residuals:                     160   BIC:                            -230.7
Df Model:                           4
Covariance Type:            nonrobust
===================================================================================================
                                      coef    std err          t      P>|t|      [0.025      0.975]
---------------------------------------------------------------------------------------------------
Intercept                           0.2052      0.015     14.083      0.000       0.176       0.234
C(incumbency)[T.1]                  0.1390      0.057      2.427      0.016       0.026       0.252
view_ratio                          0.0840      0.041      2.048      0.042       0.003       0.165
receipt_ratio                       0.3881      0.050      7.762      0.000       0.289       0.487
C(incumbency)[T.1]:receipt_ratio   -0.1874      0.080     -2.336      0.021      -0.346      -0.029
==============================================================================
Omnibus:                        0.859   Durbin-Watson:                   2.423
Prob(Omnibus):                  0.651   Jarque-Bera (JB):                0.848
Skew:                          -0.171   Prob(JB):                        0.654
Kurtosis:                       2.918   Cond. No.                         14.5
==============================================================================
```

**Appendix G.** *Average media mentions of challengers and incumbents throughout election cycle*

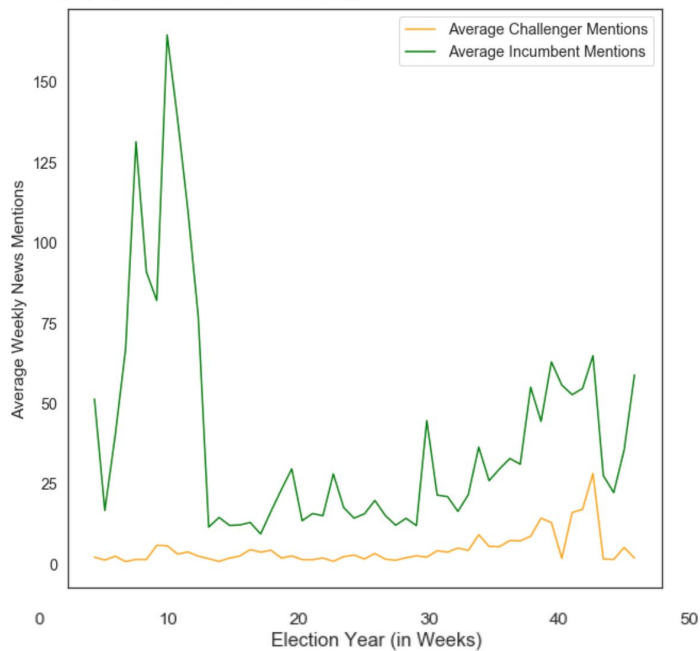

**Appendix H.** *Binomial logistic regression of multiple categorical variables (top) and probability of outcomes (below)*

```
                           Logit Regression Results
==============================================================================
Dep. Variable:                 winLose   No. Observations:                 1540
Model:                           Logit   Df Residuals:                     1534
Method:                            MLE   Df Model:                            5
Date:                 Tue, 07 Jul 2020   Pseudo R-squ.:                  0.7322
Time:                         14:27:57   Log-Likelihood:                -285.85
converged:                        True   LL-Null:                       -1067.3
Covariance Type:             nonrobust   LLR p-value:                     0.000
==============================================================================
                 coef    std err          z      P>|z|      [0.025      0.975]
------------------------------------------------------------------------------
Intercept     -1.6264      0.342     -4.753      0.000      -2.297      -0.956
challenger    -2.2321      0.336     -6.640      0.000      -2.891      -1.573
open_seat      0.7726      0.290      2.666      0.008       0.205       1.341
stronghold     2.9877      0.255     11.724      0.000       2.488       3.487
via_win        1.9793      0.228      8.666      0.000       1.532       2.427
view_win       1.0037      0.248      4.043      0.000       0.517       1.490
==============================================================================
```